\documentclass[runningheads,fleqn]{svmult}

\usepackage{makeidx}   
\usepackage{graphicx}  
\usepackage{subeqnar}  
\usepackage{multicol}  
\usepackage{physmult}  
\makeindex             

\begin{document}
\title*{Quantum state protection \\
   using all-optical feedback}
\toctitle{Quantum state protection
\protect\newline using all-optical feedback}
%
%
\titlerunning{Quantum state protection}
%
\author{Paolo Tombesi, 
Vittorio Giovannetti,
and David Vitali \\
{\itshape Dipartimento di Matematica e Fisica, and Unit\`a INFM,
Universit\`a di
Camerino, via Madonna delle Carceri, I-62032 Camerino, Italy}}

\authorrunning{Paolo Tombesi et al.}
%
%

\maketitle              
\noindent
{\bf Abstract}. An all-optical feedback scheme in which the output
of a cavity mode is used to influence the dynamics
of another cavity mode is considered. We show that 
under ideal conditions, perfect preservation against
decoherence of a generic
quantum state of the source mode can be achieved.

\section{Introduction}

Electromagnetic fields in cavities have already been 
used for quantum information
processing. For example, one of the first experimental
demonstration of a quantum gate has been implemented using two
cavity modes characterized by a nonlinear dispersive interaction
mediated by a beam of Cs atoms \cite{turchette}.
In particular, quantum information can be stored in high-Q 
electromagnetic cavities, and with this respect it is important
to develop schemes able to increase the quantum information
storage time as much as possible, and provide therefore
a good {\em quantum state protection} against the 
effects of the decoherence due to cavity leakage.

We have developed schemes for the preservation
of generic quantum states in cavities based on
feedback loops originating from
homodyne measurements \cite{goe,vitto}, or associated
with direct photodetection supplemented with
the injection of an appropriately
prepared atom \cite{prl,pra}. These schemes
provide a significative increase of the decoherence
time of an initially prepared quantum state, but are
both characterized by some limitations. In the case of 
homodyne-mediated feedback, the scheme is ``anisotropic''
in phase-space \cite{goe}, that is, it does not protect 
all the quantum states in
the same way. This is due to the fact that the dynamics in the 
presence of feedback has a privileged
direction, coinciding with that of the measured quadrature.
In the case of photodetection-mediated feedback,
the scheme is isotropic but it is affected by phase diffusion, which,
although very slowly, leads to destruction of quantum phases 
\cite{pra}.

The schemes considered in \cite{goe,vitto,prl,pra}
employ the usual implementation of optical feedback, i.e., 
electro-optical feedback,
in which the light exiting the cavity enters a detector
and the photocurrent produced is used to control the cavity 
dynamics by some electro-optical device. 
Here we show that a promising way to obtain perfect quantum
state protection, that is, the preservation of an
initially prepared quantum state for an arbitarily
large time, can be obtained by using an {\em
all-optical feedback} scheme. 
In these schemes, the output light is not detected,
but it is reflected around a feedback loop and sent into another
cavity (the driven cavity) which is coupled to the
first in some way. This scheme is an actual feedback scheme
if the loop is one-way, i.e., it goes from the source to the
driven cavity and it cannot go backward.
This can be achieved by inserting in the loop a 
system analogous to a Faraday isolator. 
With this respect, all-optical feedback schemes are an example
of cascaded quantum systems, introduced and described by
Gardiner \cite{gard} and Carmichael \cite{carm}. In these systems,
the output
from a source mode is used as an input for a second mode.
The new feature introduced by feedback is the presence of an
interaction term between the two modes, so that the source mode
dynamics is affected by the driven mode.

All-optical feedback schemes have been already studied by Wiseman and
Milburn in \cite{wisemil}. However they focus their attention
to the adiabatic regime, where the linewidth of the driven cavity
is much larger than that of the source mode, so that the driven mode 
can be adiabatically eliminated. 
In this case, an all-optical feedback scheme reduces 
to an analogous electro-optical feedback scheme whenever the 
interaction between driven and source mode has a quantum non demolition
(QND)-like form, that is, it is a product of source and driven mode
operators. In this case, in fact, the role of the driven mode is 
completely equivalent to that of a detection apparatus
\cite{wisemil}. On the contrary, all-optical feedback cannot be reduced to an
electro-optical analogous in the case of a non-factorized form
of interaction Hamiltonian. This is the
most interesting case and 
in this work we shall only consider this case, 
which can be experimentally realized,
for example,
using a simple set up involving a single cavity. 
In this case, one polarization mode plays the role of the source mode
and an orthogonal polarization mode plays the role of the driven
system. The unidirectional coupling is provided by an optically active
element supplemented with two polarized beam splitter and a polarizer. 
We shall see that the scheme is able to provide an
``isotropic'', i.e. phase-independent, quantum state protection for 
the source mode.
More interestingly, we show that in the ideal limit of
unit efficiency of the feedback loop, feedback parameters can be 
chosen so to achieve perfect state protection, i.e., 
perfect freezing of the source mode dynamics.

\section{The all-optical feedback scheme}

Let us briefly recall the theory of cascaded quantum systems 
developed by Gardiner and Carmichael in \cite{gard,carm} and 
reconsidered by Wiseman and Milburn in \cite{wisemil}. 
This theory describes two systems, the source system and the driven 
system, which are unidirectionally coupled. This broken symmetry can 
be naturally obtained in optical systems when the coupling is realized
by a reservoir of electromagnetic waves traveling in one direction.
Experimentally this one-way isolation can be obtained using a Faraday 
rotator. 
This means that the source emits photons influencing the dynamics of 
the driven system, while the radiation emitted by the driven system
does not affect the source. The source and the driven system can be 
generic quantum system, but here we shall consider the case of two
optical cavities.
If we denote with $a_{1}$ and $\gamma_{1}'$ the annihiliation 
operator and the decay rate of the source cavity mode, and with
$a_{2}$ and $\gamma_{2}'$ the corresponding quantities for the driven 
cavity mode, the dynamics of a generic operator $c(t)$ 
can be obtained using the input-output
theory \cite{qnoise}, yielding the following quantum Langevin equation 
\cite{gard}:
\begin{eqnarray}
&&\dot{c}(t)= -\frac{i}{\hbar}\left[c(t),H\right]  
-\left[c(t),a_{1}^{\dagger}(t)\right]\left\{\frac{\gamma_{1}'}{2}a_{1}(t)+
\sqrt{\gamma_{1}'}a_{in}(t)\right\} \nonumber  \\
&&+
\left\{\frac{\gamma_{1}'}{2}a_{1}^{\dagger}(t)+
\sqrt{\gamma_{1}'}a_{in}^{\dagger}(t)\right\}
\left[c(t),a_{1}(t)\right] \label{inpu} \\
&& 
-\left[c(t),a_{2}^{\dagger}(t)\right]\left\{\frac{\gamma_{1}'}{2}a_{2}(t)+
\sqrt{\gamma_{1}'\gamma_{2}'}a_{1}(t-\tau)+
\sqrt{\gamma_{2}'}a_{in}(t-\tau)\right\} \nonumber \\
&&+
\left\{\frac{\gamma_{2}'}{2}a_{2}^{\dagger}(t)+
\sqrt{\gamma_{1}'\gamma_{2}'}a_{1}^{\dagger}(t-\tau)
+\sqrt{\gamma_{2}'}a_{in}^{\dagger}(t-\tau)\right\}
\left[c(t),a_{2}(t)\right] \nonumber \;.
\end{eqnarray}
We have considered the presence of a total system Hamiltonian $H$; 
then $a_{in}(t)$ is the input noise at the source cavity, with
$\left[a_{in}(t),a_{in}(t')\right]= \delta(t-t')$, and $\tau$ is such that
$c\tau$ is the distance between the two cavities.
When $H=H_{1}+H_{2}$ is the sum of a source Hamiltonian $H_{1}$
and a driven mode Hamiltonian $H_{2}$, we have a cascaded system and
the meaning of (\ref{inpu}) is evident. The equation of an operator 
of the source cavity does not involve the last two lines of (\ref{inpu}),
and one has the usual quantum Langevin for the source cavity, since 
the driven cavity has no effect on it. On the contrary, in the case 
of a driven cavity operator, the second and third term of the right 
hand side of 
(\ref{inpu}) is zero and one has the usual quantum Langevin equation
but with an input field equal to the output field from the source 
cavity, delayed by $\tau$.
In the case of cascaded systems, the delay $\tau$ is an arbitrary 
constant, which is essentially irrelevant for the physics of the 
problem. In fact, the results for a given value of the delay $\tau$
can be obtained from those with another value for $\tau$ with simple,
appropriate, adjustments. It is evident that the easiest case is the
limiting case of a vanishingly small delay $\tau \rightarrow 0$, which 
involves the input noise at time $t$, $a_{in}(t)$, only, and this explains why 
the zero delay case is usually considered.

The delay $\tau$ becomes an important physical parameter in the 
presence of some feedback process, i.e., when the driven mode can 
affect in some way the source mode dynamics. This could be done, 
for example, simply
by removing the Faraday isolation, i.e., restoring the inversion 
symmetry, but this simply means going back to the trivial case of 
two interacting 
systems. A more interesting situation is obtained when the 
unidirectional coupling is left unchanged, and feedback 
from the driven to the source system is obtained through a coupling 
Hamiltonian term. This means that the Langevin equation (\ref{inpu}) 
is still valid, but with a non-decomposable total system Hamiltonian
$ H = H_{1}+H_{2}+H_{int} $,
so that the two cavity modes are no more real cascaded systems.
The presence of the interaction term $H_{int}$ implies that the two 
cavities have to overlap spatially, at least partially. In this case 
one essentially realizes an {\em all-optical feedback scheme}, 
because in this way one tries to implement a control of the source 
mode dynamics through an optical loop involving the driven cavity 
and its interaction with the source mode.
In this case, the delay $\tau$ acquires the meaning of a 
feedback loop transit time and the $\tau \neq 0$ case now corresponds
to a truly non-Markovian dynamics \cite{vitto}. The Markovian 
limiting case $\tau \rightarrow 0$ becomes now a well specified 
physical assumption, which is justified only in the case when the
feedback delay $\tau$ is much smaller than the typical timescale of 
the dynamics of the system of interest, i.e., of the source mode. 
Since we are concerned with the preservation of a generic quantum 
state generated in the source cavity, the relevant timescale here is 
the decoherence time, which is given by 
$t_{dec} \simeq (\gamma_{1}' \bar{n})^{-1}$, where $\bar{n}$ is the 
mean number of photons \cite{milwal}. The feedback loop delay 
time is instead of the order of a single cavity transit time
$\tau \simeq 2L/c$ ($L$ is the cavity length) and since 
$1/\gamma_{1}' = 2L/cT$, where $T$ is the cavity mirror 
transmittivity, it is evident that for good cavities, the Markovian 
limit $\tau \rightarrow 0$ can be safely assumed even for quantum states
of the source mode with a quite large number of photons.

In the Markovian limit $\tau \rightarrow 0$, the quantum Langevin equation 
(\ref{inpu}) becomes equivalent to a master equation for the joint density 
matrix $D(t)$ of the source and driven modes. We consider the 
most common case of a vacuum reservoir, that is, $\langle a_{in}(t)
a_{in}^{\dagger}(t)\rangle =\delta (t-t')$ (the case of
more general input white noises is considered in \cite{wisemil}). 
Moreover we generalize to the realistic situation in which the losses 
in each cavity are not due only to coupling with the vacuum 
electromagnetic modes responsible for the unidirectional coupling 
between the source and the driven mode (with rates $\gamma_{i}'$), 
but also to the coupling with some other 
unwanted modes (absorption and diffraction losses), with rates 
$\eta_{i}$. The general master equation for all-optical feedback
in the $\tau \rightarrow 0$ limit is therefore
\begin{eqnarray}
\label{maste}
&&\dot{D} = -\frac{i}{\hbar}\left[H,D\right]
+\frac{\gamma_{1}'+\eta_{1}}{2}\left(
2a_{1}D a_{1}^{\dagger} - a_{1}^{\dagger}a_{1}D -
D a_{1}^{\dagger}a_{1}\right)  \\
&& +\frac{\gamma_{2}'+\eta_{2}}{2}\left(
2a_{2}D a_{2}^{\dagger} - a_{2}^{\dagger}a_{2}D -
D a_{2}^{\dagger}a_{2}\right) \nonumber \\
&&+\sqrt{\gamma_{1}'\gamma_{2}'}\left\{\left[a_{1}D,a_{2}^{\dagger}\right]
+\left[a_{2},D a_{1}^{\dagger}\right]\right\} \nonumber \;.
\end{eqnarray}
In this work we apply this master equation to a set up which could be
realized experimentally in a quite straightforward way and which is
schematically shown in 
Fig.~1. The source and the driven cavity coincide and the two 
annihilation operators $a_{1}$ and $a_{2}$ describe two frequency 
degenerate, orthogonally polarized modes of the cavity. As discussed 
in detail in \cite{wisemil}, in order to have a feedback scheme with no 
electro-optical analog, one has to choose an interaction Hamiltonian 
$H_{int}$ which cannot be factorized into a source and a driven term.
We choose the simplest case, a mode conversion term, which can be 
realized even without a nonlinear medium, but with a simple half-wave 
plate, i.e., a polarization rotator. In the frame rotating at the 
frequency of the modes, one has
\begin{equation}
\label{hi}
H= i\hbar \frac{g \sqrt{\gamma_{1}'\gamma_{2}'}}{2} 
\left(a_{1}^{\dagger}a_{2}-a_{2}^{\dagger}a_{1}\right) \;,
\end{equation}
where we have defined the coupling in terms of the dimensionless
constant $g$. In this case
the unidirectional coupling can be simply realized using two 
polarized beam splitters, a Faraday rotator and an half-wave plate 
(see Fig.~1).

\begin{figure}
\centerline{\includegraphics[width=.7\textwidth]{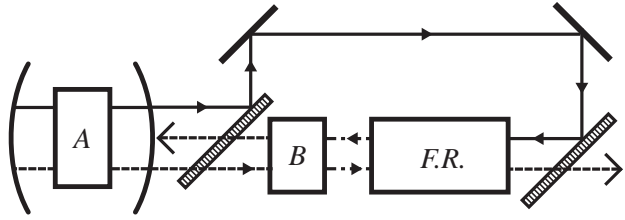}}
\caption[]{Scheme of the proposed all-optical feedback loop. The 
source mode (full line) and the driven mode (dashed line) are,
respectively, horizontally and vertically polarized and are coupled 
within the cavity by the half-wave plate {\em A}.
The source mode passes through an optically active element {\em F.R.}
and the half-wave plate {\em B}, both 
rotating its polarization by $\pi/4$ radians, so that it finally
drives the driven mode in the cavity. The output from the driven mode
cannot come back into the cavity because of the action of the
optically active element and of the polarized beam splitter.}
\label{fig.1}
\end{figure}

\section{The dynamics of the system}

Before studying the dynamics of the two coupled cavity modes, it is 
convenient to consider the adiabatic regime where the driven mode 
bandwidth $\gamma_{2}=\gamma_{2}'+\eta_{2}$ is much larger than that
of the source mode. This limit will show in which way the 
optical feedback loop is able to inhibit the decohering effects of 
photon leakage. When $\gamma_{2}$ is much larger than the other 
parameters, the driven mode can be adiabatically eliminated so to get 
a master equation for the reduced density matrix of the source mode 
alone $\varrho$. The driven mode will always be very close to the vacuum 
state, so that we can expand the total density matrix $D$ as
\begin{eqnarray}
&& D=w_{0} \otimes|0\rangle\langle 0|
+w_{1}\otimes |1\rangle\langle 0|+w_{1}^{\dagger}\otimes 
|0\rangle\langle 1| \nonumber \\
&&+ w_2' \otimes |1\rangle \langle 1|
+w_2\otimes|2\rangle\langle 0|
+w_2^{\dagger} \otimes|0\rangle\langle 2| \,,
\label{Dofrho2}
\end{eqnarray}
where $|n\rangle $, $n=0,1,2$, are the lowest driven mode Fock states.
Inserting this expression in the master equation (\ref{maste}), one 
gets a set of coupled equations for $w_{i}$, which, at lowest order 
in $\gamma_{2}^{-1}$, yields the following master equation for
the source mode reduced density matrix $\varrho =w_{0}+w_{2}'$
\begin{equation}
	\dot{\varrho}(t)=\frac{\gamma_{1}}{2}\left[1+g(2+g)\frac{\gamma_{1}'
	\gamma_{2}'}{\gamma_{1}\gamma_{2}}\right]\left(
	2a_{1}\varrho a_{1}^{\dagger} - a_{1}^{\dagger}a_{1}\varrho -
\varrho a_{1}^{\dagger}a_{1}\right)\;,
	\label{reduma}
\end{equation}
where $\gamma_{1}=\gamma_{1}'+\eta_{1}$ is the total decay rate of 
the source mode. Equation~(\ref{reduma}) shows that, in the adiabatic limit, 
the dynamics of the source mode in the presence of the optical 
feedback loop is still described by the standard 
vacuum optical master equation, but with a {\em renormalized decay rate}
$ \gamma_{1}^{eff}=
\gamma_{1}\left[1+g(2+g)\eta ^{2}\right]$, where we have defined
the feedback efficiency $\eta = \sqrt{\gamma_{1}'
\gamma_{2}'/\gamma_{1}\gamma_{2}}$ (a decay rate 
renormalization in the adiabatic limit is already predicted  
in \cite{wisemil}). It is easy to see that the 
feedback is optimal, i.e., the effective decay rate $\gamma_{1}^{eff}$
is minimized, when the dimensionless mode conversion coupling $g=-1$, 
and in this case $\gamma_{1}^{eff}= \gamma_{1}\left[1-\eta ^{2}\right]$.
Therefore in the ideal limit of perfect feedback $\eta =1$
(i.e., no light is lost in the loop due to diffraction or absorption), 
when $g=-1$ and $\gamma_{2} \gg \gamma_{1}$, all-optical feedback 
completely freezes the source mode dynamics, that is,
it realizes perfect preservation of an initial quantum state.
In this ideal case, the whole source mode output is collected and 
converted by the all-optical loop into driven mode light, which is then
efficiently converted again within the cavity into source mode light.
No source mode photon is lost in the loop, and more importantly, 
when $g=-1$, optical feedback acts {\em in phase}, yielding a complete
suppression of photon leakage. This phase-sensitive aspect of 
all-optical feedback cannot be achieved with electro-optical feedback.
For example in \cite{pra}, we have studied a 
direct-photodetection based electro-optical feedback loop, feeding 
back a photon in the cavity through atomic injection, whenever a 
photon is lost and detected. In this case, perfect state preservation 
is not achieved even in the ideal limit of unit feedback efficiency, 
because the fed back photon has no phase relationship with those in 
the cavity, and one is left with an unavoidable, even though slow, 
phase diffusion.
This study of the adiabatic limit $\gamma_{1}/\gamma_{2} \ll 1$ shows 
that, with all-optical feedback, perfect state preservation is in 
principle possible using the scheme of Fig.~1. We now study the exact 
dynamics of the two coupled modes by solving the master equation 
(\ref{maste}) in order to see the performance of the scheme as a 
function of the feedback efficiency $\eta $ and of the adiabaticity
parameter $\gamma_{1}/\gamma_{2}$. 
We shall consider a 
factorized initial condition $D(0)=\varrho_{1}(0) \otimes \varrho_{2}(0)$.
It is convenient to expand the initial conditions $\varrho_{j}(0)$ 
using the $R$ representation \cite{qnoise}
\begin{equation}  
\varrho_{j}(0)= \frac{1}{\pi^{2}}\int \D^{2}\alpha_{j}\D^{2}\beta_{j} 
\; R_{j}(\beta_{j}^{*},\alpha_{j}) \;| \beta_{j} \rangle \langle \alpha_{j}| \; 
e^{-(|\alpha_{j}|^{2}+|\beta_{j}|^{2})/2} \; ,
\label{condin}
\end{equation}
where $|\alpha_{j} \rangle$ and $| \beta_{j} \rangle$ are coherent 
states and
\begin{equation}  
R_{j}(\beta_{j}^{*},\alpha_{j}) = \langle \beta_{j}| \varrho_{j}(0) 
|\alpha_{j}\rangle \; e^{(|\alpha_{j}|^{2}+|\beta_{j}|^{2})/2}
\label{erre}
\end{equation} 
is the $R$ function, analytic in the two complex variables
$\alpha_j$ and $\beta_{j}^{*}$.
We are interested in the dynamics of the source mode only
and, even though we do not adiabatically eliminate
the driven mode, we shall always consider it more damped than the source mode.
It is therefore reasonable to assume an initial vacuum state
for the driven mode $\varrho_2(0) =|0\rangle \langle 0 |$, which 
means $R_{2}(\beta_{2}^{*},\alpha_{2}) = 1$. Moreover
we shall trace over the driven mode and focus on the reduced Wigner
function of the source mode $
W_{1}(\alpha,\alpha^{*},t) = \int \D^{2} \beta \, 
W(\alpha,\alpha^{*},\beta,\beta^{*},t) $.
The time evolution of this reduced Wigner function can be 
determined exactly in terms of the $R$ representation of the source mode
initial condition, and after some gaussian integrations one gets
\begin{eqnarray}
W_{1}(\alpha,\alpha^{*},t) &=&
\frac{2}{\pi^{3}}\int \D^{2}\alpha_{1}  
\D^{2}\beta_{1} \;
R_{1}(\beta_{1}^{*},\alpha_{1}) \; 
e^{-|\alpha_{1}|^{2}-|\beta_{1}|^{2} + \beta_{1} \alpha_{1}^{*}} 
\times \nonumber \\
& & \times \;  \exp \Big[ 2 \big( \alpha - {\cal F}(t) \beta_{1}\big) 
\big( {\cal F}(t) 
\alpha_{1}^{*} -\alpha^{*}\big) \Big] \; ,
\label{sol}
\end{eqnarray}
where 
\begin{equation}
{\cal F}(t)= \frac{m+\gamma_{1} - 
\gamma_{2}}{2m}\E^{-\left(\gamma_{1}  +
\gamma_{2}+m \right)t/4}
+\frac{m-\gamma_{1} + 
\gamma_{2}}{2m} 
\E^{-\left(\gamma_{1}  +
\gamma_{2}-m \right)t/4}
\label{funzione}
\end{equation}
and $
m=\sqrt{ (\gamma_{1} - \gamma_{2})^{2} - 4 \gamma_{1} 
\gamma_{2} \eta^2 (2 + g ) g} $.
It is interesting to notice that the exact dynamics
of the source mode is completely characterized by the function ${\cal F}(t)$
of (\ref{funzione}) only. It can be shown that ${\cal F}(t)$
is always a nonincreasing function of time. In particular, in
the absence of feedback
($g=0$) one has ${\cal F}(t) = \exp\left(-\gamma_1 t/2\right)$, since
the source mode is not affected by the driven mode. In the adiabatic limit,
one can see from (\ref{funzione}) that, at first order in $\gamma_1/\gamma_2$,
one has ${\cal F}(t) = \exp\left(-\gamma_1 \left(1+\eta^2 (2+g)g \right)
t/2\right)$, 
implying (see also section II), that in the ideal case
$\eta =1$ and $g=-1$, it is ${\cal F}(t) =1$ and therefore the source mode 
dynamics is completely frozen.

It is instructive to apply the general expression of the time evolved Wigner 
function of (\ref{sol}) to some specific initial states of the source mode.
The paradygm case for decoherence studies is the Schr\"odinger cat case,
$\varrho_1(0) =|\psi(0) \rangle \langle \psi(0) |$, with
\begin{equation}
|\psi (0)\rangle = \frac{1}{\sqrt{2 (1 + e^{-2 
|\alpha_{0}|^{2}}\cos \varphi)}} \big( | \alpha_{0} \rangle + e^{i\varphi}
| - \alpha_{0}  \rangle \big) \;;
\label{gatto}
\end{equation}
applying (\ref{sol}) one gets that the corresponding time evolution of the
Wigner function is 
\begin{eqnarray}
W_{1}(\vec{x},t) &=& \frac{1}{\pi (1 + 
e^{-2 |\vec{\alpha_{0}}|^{2}}\cos \varphi)}\Big( 
e^{-2|\vec{x} - \vec{\alpha_{0}}{\cal F}(t)|^{2}} +
e^{-2|\vec{x} + \vec{\alpha_{0}}{\cal F}(t)|^{2}} \nonumber \\
& +& 2 e^{-2 \big( |\vec{x}|^{2} - |\vec{\alpha_{0}}{\cal F}(t)|^{2} +
 |\vec{\alpha_{0}}|^{2}\big) } \cos \Big(4 ( \vec{x} \wedge 
 \vec{\alpha_{0}} ) \cdot \hat{z}- \varphi \Big) \Big) \;,
\label{wignergatto}
\end{eqnarray}
where $\vec{x}=\left({\rm Re}\{ \alpha \}, {\rm Im}\{ \alpha \},0\right)$,
$\vec{\alpha_0}=\left({\rm Re}\{ \alpha_0 \}, {\rm Im}\{ \alpha_0 \},0\right)$ 
and $\hat{z}=(0,0,1)$. From (\ref{wignergatto}) one can see
the isotropic properties of the all-optical feedback scheme
studied here, since the state of the source mode
depends upon the 
angle between $\vec{x}$ and $\vec{\alpha_{0}}$ only.
The time evolution of a Schrodinger cat state 
with $\alpha_0 =2\I$ and $\varphi =0$ is displayed in Fig.~2:
in (a) the initial condition is shown, while in (b) and in (c)
the state evolved in the presence of feedback after two decoherence
times $t_{dec} =1/(2\gamma_1 |\alpha_0|^2)$ and $20t_{dec}$ are
respectively shown. In (d) the state evolved after $2t_{dec}$ in the {\itshape 
absence} of feedback is instead shown. What is relevant is that, with 
achievable feedback parameters $g=-1$, $\eta = 0.95$, and $\gamma_1/\gamma_2
=10^{-3}$, one gets a very good preservation of the initial mesoscopic
Schr\"odinger cat state ($\bar{n}=4$) after two decoherence times.
With all-optical feedback, one has a decohered cat state similar to
that obtained in the absence of feedback after $2t_{dec}$, only after 20
decoherence times.

\begin{figure}
\centerline{\includegraphics[width=.78\textwidth]{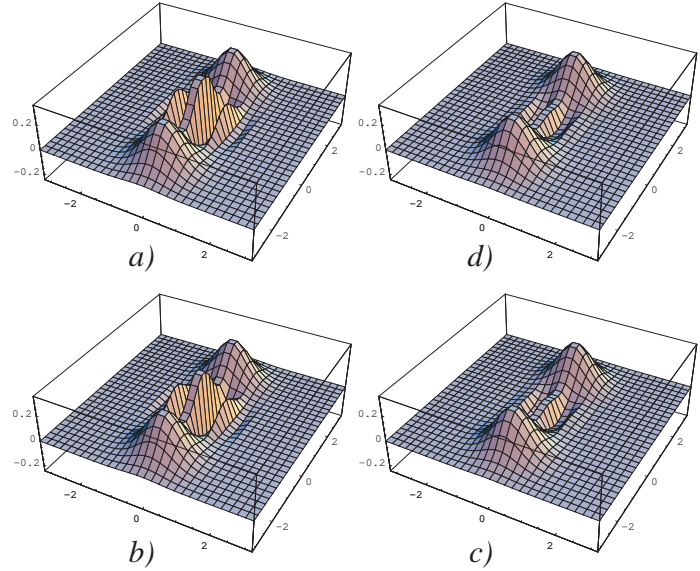}}
\caption[]{Time evolution of the source mode
Wigner function of the cat state of
equation (\protect\ref{gatto}) with $\alpha_0=2\I$, $\varphi=0$. (a) 
Initial Wigner function; (b) state after two decoherence times $t_{dec}$
in the presence of feedback; (c) state after $t = 20t_{dec}$
in the presence of feedback. Parameter values are
$g=-1$, $\eta=0.95$, and $\gamma_1/\gamma_2=10^{-3}$. 
(d) shows the state after $t=2t_{dec}$
in the absence of feedback ($g=0$).}
\label{fig.2}
\end{figure}

Similar qualitative results are obtained with a different initial
pure quantum state of the source mode, i.e., the linear superposition of
Fock states $
\psi(0) \rangle = \left(|2\rangle +\sqrt{2} |4\rangle\right)/\sqrt{3} $.
The time evolution of the Wigner function is shown in Fig.~3, where,
again, (a) shows the initial state, (b) and (c) show
the state evolved in the presence of feedback after $2t_{dec}$ 
and after $20t_{dec}$ respectively, and
(d) shows the state evolved after $2t_{dec}$ in the 
absence of feedback. The feedback parameters are the same as in Fig.~2.
One has again a very good preservation
of quantum coherence after two decoherence times.

\begin{figure}
\centerline{\includegraphics[width=.78\textwidth]{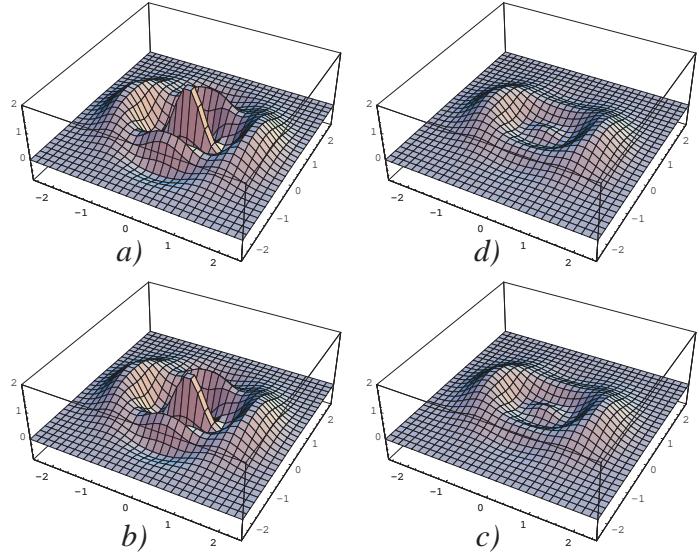}}
\caption[]{Time evolution of the source mode
Wigner function of the linear superposition of Fock states
$
|\psi(0) \rangle = \left(|2\rangle +\sqrt{2} |4\rangle\right)/\sqrt{3} $. (a) 
Initial Wigner function; (b) state after two decoherence times $t_{dec}$
in the presence of feedback; (c) state after $t = 20t_{dec}$
in the presence of feedback; (d) state after $t=2t_{dec}$
in the absence of feedback. Parameter values are the same as in Fig.~2.}
\label{fig.3}
\end{figure}

A more quantitative characterization of the preservation properties
of the all-optical feedback scheme is obtained from the study of the
fidelity of the initial state,
$
F(t) = {\rm Tr} \left\{ \varrho_{1}(t) \varrho_{1}(0) \right\} $.
Using (\ref{sol}) it is possible to write the fidelity of a generic
initial state in terms of its $R$ representation (\ref{erre})
and the function ${\cal F}(t)$ as
\begin{eqnarray}
\label{fidelio}
F(t) &=& \frac{1}{\pi^{2}} \int  \D^{2}\alpha_{1} 
\D^{2}\beta_{1} e^{-|\alpha_{1}|^{2} -|\beta_{1}|^{2}} \\ 
&\times & R_{1}(\beta_{1}^{*},(1-{\cal F}^{2}(t) \beta_{1} + {\cal F}(t) \alpha_{1})
R_{1}(\alpha_{1}^{*},{\cal F}(t) \beta_{1}) \;. \nonumber
\end{eqnarray}
The time evolution of the fidelity in the case of the initial Schr\"odinger
cat state of (\ref{gatto}) is shown in Fig.~4. In (a) $F(t)$ is plotted
for different values of the feedback efficiency $\eta$ and with fixed values 
for the coupling constant $g$ (the optimal choice $g=-1$ is considered)
and for the ratio $\gamma_1/\gamma_2$. As expected, the preservation of the
quantum state worsens for decreasing efficiencies. In (b) the effect of the 
timescale separation between source and driven mode is studied and $F(t)$
is plotted for different values of $\gamma_1/\gamma_2$ at fixed values
for the coupling and the feedback efficiency (the optimal values $g=-1$ and
$\eta =1$ are considered). We notice in particular
that the decay rates ratio $\gamma_1/\gamma_2$
plays an important role and that only in the adiabatic limit 
$\gamma_1/\gamma_2 \ll 1$ one gets a fidelity very close to one. 
Even in the adiabatic regime and in the
ideal case (see for example the curve corresponding
to $\gamma_1/\gamma_2 = 10^{-3}$), a finite value for $\gamma_2$ determines
an appreciable initial slip from the condition $F(t)=1$ at small times,
before the fidelity saturates to its asymptotic value.

\begin{figure}
\centerline{\includegraphics[width=.8\textwidth]{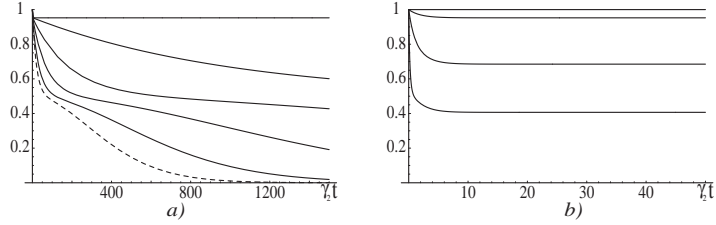}}
\caption[]{Time evolution of the fidelity $F(t)$
in the case of an initial Schr\"odinger cat state (see
(\protect\ref{gatto}) with $\alpha_0=5$ and $\varphi=0$). (a) 
shows $F(t)$ for different values of the feedback efficiency $\eta$,
with $g=-1$ and $\gamma_1/\gamma_2=10^{-3}$. 
From the top to the bottom: $\eta =1$, $\eta =0.99$, $\eta=0.97$,
$\eta =0.95$, $\eta =0.90$; the dashed line refers to $g=0$ (absence of
feedback).
(b) shows $F(t)$ for different values of the decay rates ratio
$\gamma_1/\gamma_2$ with $g=-1$ and $\eta=1$ kept fixed. From top
to bottom: $\gamma_1/\gamma_2= 0$, $\gamma_1/\gamma_2=10^{-3}$, 
$\gamma_1/\gamma_2=10^{-2}$, $\gamma_1/\gamma_2=10^{-1}$.}
\label{fig.4}
\end{figure}
We have studied the behavior of the fidelity for a large class
of initial states, as for example the linear superposition of Fock states
of Fig.~3, and we have always found a behavior completely analogous
to that shown in Fig.~4.

In conclusion, we have proposed an all-optical feedback scheme
involving two orthogonally polarized modes in a cavity. The output 
light from the source mode is sent back using a Faraday isolator into
the other, driven, mode and feedback is achieved by coupling the
two modes within the cavity via a half-wave plate. In the adiabatic limit
in which the driven mode is much more damped than the source mode,
it is possible to choose the coupling constant so that in the ideal case
of unit feedback efficiency one has freezing of the source mode dynamics, 
and therefore perfect preservation of quantum coherence.
We have also shown that the protection capabilities of the scheme
remain good even in the case of realistic values of the feedback efficiency.

%

\end{document}